\documentclass[preprint,showpacs,preprintnumbers,amsmath,amssymb]{revtex4}

\usepackage{graphicx}
\usepackage{dcolumn}
\usepackage{bm}

\begin{document}

\preprint{APS/123-QED}

\title{Stochastic acceleration by multi-island contraction during turbulent
 magnetic reconnection}

\author{Nicolas H. Bian}
\author{Eduard P. Kontar}%
\affiliation{School of Physics \& Astronomy, The University, Glasgow
G12 8QQ, Scotland, UK}

\date{\today}

\begin{abstract}
The acceleration of charged particles in magnetized plasmas is
considered during turbulent multi-island magnetic reconnection. The
particle acceleration model is constructed for an ensemble of
islands which produce adiabatic compression of the particles. The
model takes into account the statistical fluctuations in the
compression rate experienced by the particles during their transport
in the acceleration region. The evolution of the particle
distribution function is described as a simultaneous first and
second-order Fermi acceleration process. While the efficiency of the first-order process
is controlled by the average rate of compression, the second order process involves the variance
in the compression rate. Moreover, the acceleration efficiency
associated with the second-order process involves both the Eulerian
properties of the compression field and the Lagrangian properties of
the particles. The stochastic contribution to the acceleration is
non-resonant and can dominate the systematic part in the case of a
large variance in the compression rate. The model addresses the role
of the second-order process, how the latter can be related to the
large-scale turbulent transport of particles and explains some
features of the numerical simulations of particle acceleration by
multi-island contraction during magnetic reconnection.
\end{abstract}

 \maketitle

\section{Introduction}
The production of non-thermal particles in magnetized plasmas is an
ubiquitous complex phenomenon which is believed to involve also
magnetic reconnection. Magnetic reconnection is the process that
controls the conversion of magnetic energy into kinetic
energy\cite{RevModPhys.82.603}; it is the driver of impulsive
phenomena such as solar flares, substorms in the Earth's
magnetosphere and disruptions in laboratory fusion devices. The
relation between magnetic reconnection and particle acceleration has
been extensively discussed in the terrestrial magnetosphere based on
in-situ observations \cite{2012SSRv..tmp...20B, 2002PhRvL..89s5001O,2008NatPh...4...19C}.
Moreover, X-ray observations and studies of the
energy budget during solar flares indicate that a significant
fraction of the magnetic energy released in a flare is carried by
the accelerated $10-100$ keV non-thermal electrons \cite{2011SSRv..159..107H}.
However, how particles can be accelerated in large numbers to high energies as
the magnetic field lines reconnect remain an outstanding problem.

Numerical PIC simulations aiming to address the problem of particle
acceleration during magnetic reconnection in a self-consistent
manner, have confirmed that particles are efficiently accelerated in
the vicinity of the X-line by reconnection electric fields
\cite{2001JGR...10625979H, 2006GeoRL..3313104P}. However, an
important limitation of X-type acceleration mechanisms is that they
hardly explain alone the large number of accelerated particles, in
particular during a solar flare, because the volume occupied by a
current sheet where the strong electric
field capable of particle acceleration is present,
is quite small.

Hence, intensive efforts have been made to understand the role
played by the region inside the separatrices for particle
acceleration, leaning toward the idea of O-type acceleration
mechanisms which take advantage of the closed geometry of the
field within magnetic islands. Drake et
al.\cite{2006Natur.443..553D} have developed a model of particle
acceleration which is based on the dynamical motion of the islands.
They show that particles trapped in the contracting magnetic field
of the islands are adiabatically compressed and therefore can be
efficiently accelerated through a first-order Fermi process. In
addition, many studies have revealed the importance of
magnetohydrodynamics turbulence
\cite{1984PhRvL..53.1449M,1994ApJS...90..719K,2006PhRvL..96o1102O}
and plasmoid dynamics \cite{2010ApJ...714..915O,2011PhPl...18b2903T}
with regards to particle acceleration in a reconnecting plasma.
Current sheets are naturally prone to tearing and their
fragmentation lead to the formation of magnetic islands having a
complex multiscale and intermittent dynamical behavior
\cite{2010PhRvL.105w5002U,2010PhPl...17a0702F}. Our goal in this
Letter is to study the effect on particle acceleration of the
ensemble of contracting islands and develop a simple model of
particle acceleration during turbulent magnetic reconnection.

When a magnetic island changes its length $L$ at the velocity
$V=dL/dt$ particles that are trapped within the island change their
speed $v$ according to $dv/dt=-(\alpha V/L)v\equiv Wv$. This
relation, derived in \cite{2006Natur.443..553D} is a consequence of
the conservation of the longitudinal action for particles trapped
within an island and the coefficient $\alpha$ represents the
relative magnitude of the reconnecting magnetic field, i.e.
$\alpha=(\delta B/B_{0})^{2}$. As a result, magnetic islands that
are contracting at a speed of the order of the Alfven speed $V \sim
V_{A}$, accelerate the trapped particles through adiabatic
compression, provided $v \gg V_{A}$. The acceleration rate
associated with this first-order Fermi process is given by $\alpha
V_{a}/L_{0}$,  where $L_{0}$ is the typical length of the islands.
If on the contrary the islands are expanding, then first-order
adiabatic deceleration of the particles will result at the same
energy independent rate. When contraction is magnetically favorable
and when the energy gained from the magnetic field by the particles
is balanced with energy losses, including transport losses and/or
back-reaction of the accelerated particles, power-law distribution
in particles energy may be obtained, which are determined by
standard techniques, as was originally done in
\cite{2006Natur.443..553D}.

In the case of a first-order Fermi process, the rate of energy
gained by the particles is proportional to the mean compression
$\langle W\rangle $, where the brackets $\langle \rangle $ denote an average
over the ensemble of islands in the system, possibly weighted by the relative number of
islands that are undergoing contraction
\cite{2006Natur.443..553D,2010ApJ...709..963D}. The mere existence
of this average, or the range of possible contraction rates, suggests to consider
also the effect of the finite variance in the adiabatic compression experienced
by the particles in the sea of islands. Indeed, the contraction rate changes in time
due to firehose condition \cite{2010ApJ...709..963D}, so an assemble
of contracting island will have non-zero variance.  Further, PIC simulations \cite{2010ApJ...714..915O}
emphasize the bouncing motion of merged islands, so that a contracting motion
of an island is followed by an expanding motion.

For an ensemble of multiple contracting islands, the presence
of non-zero average $ \langle W\rangle $ and non-zero
$\langle (W-\langle W\rangle )^{2}\rangle $ leads to both first and
second order accelerations. The mean controls the first order Fermi acceleration
and additional statistical acceleration occurs at a rate proportional to the variance
of the compression, also when the mean compression
rate $\langle W\rangle $ is non-zero. A continuity equation can be written
for the omnidirectional particle distribution function $F(p,t)=4\pi p^{2}f(p,t)$
\cite{2006Natur.443..553D,2010ApJ...709..963D},
\begin{equation}\label{1}
\frac{ \partial F(p,t)}{\partial t}+\frac{\partial }{\partial p}\left[\left(\frac{dp}{dt}\right)F(p,t)\right]=0,
\end{equation}
where $p$ is the particle momentum with the time rate of change in
momentum  given by
\begin{equation}
\frac{dp}{dt}=-\frac{\alpha V}{L}p\equiv Wp.
\end{equation}
In \cite{2006Natur.443..553D,2010ApJ...709..963D}, a term modeling the
effect of escape of particles out of the acceleration region is also included in
Eq.(\ref{1}).

Let us consider the case where
the compression rate is small and assume first
that it is a function of time only with zero average,
i.e. bouncing motion of the islands
\begin{equation}
\langle W(t) \rangle=0,
\end{equation}
and its correlation function decays exponentially,
\begin{equation}
C(t)=\langle W^{2}(t)\rangle \exp(-t/\tau_{c}).
\end{equation}
In this cases, even when on average the islands are neither contracting nor expanding,
i.e. $ \langle W\rangle =0 $, there is a stochastic acceleration effect that remains operative.
Although, the particles do not experience any systematic change in their energy,
the average particle energy could still grow with the acceleration efficiency associated
with a second-order Fermi process is proportional to the variance of the compression
$\langle W^{2}\rangle $.  Therefore, we obtain that the mean omnidirectional
distribution function $F_{0}(p,t)$ obeys the diffusion equation
\begin{equation}\label{d1}
\frac{\partial F_{0}(p,t)}{\partial t}=D\frac{\partial }{\partial
p}p\frac{\partial}{\partial p} p F_{0}(p,t),
\end{equation}
with the diffusion coefficient in momentum space given by
\begin{equation}\label{con}
D=\int _{0}^{\infty}dt C(t)=\tau_{c} \langle W^{2}(t)\rangle .
\end{equation}
The mean distribution function $F_{0}(p,t)$ solution of Equation (\ref{d1}) is
the normal distribution with respect to the variable
$u=\ln(p/p_{0})$. Indeed, the particle dynamics is described by the
Langevin equation $du/dt=W(t)$ with $\langle W(t)\rangle =0$.
Therefore, $F_{0}(u,t)$ satisfies the standard diffusion equation
\begin{equation}\label{diffu}
\frac{\partial F_{0}(u,t)}{\partial t}=D\frac{\partial^{2}F_{0}(u,t)
}{\partial u^{2}},
\end{equation}
which also confirms that fluctuations in the compression rate are
responsible for the growth of the variance in the momemtum
distribution function.

An account for the effect on the stochastic acceleration of the
spatial transport of particles in the pulsation field of the islands
may be given on the following basis. Let us shrink the volume of
each island into a point, this point being characterized by its
compression $W(\mathbf{x},t)$ with $\mathbf{x}$ being the position
of the center of the islands. Moreover, we envisage a situation
where the large scale spatial transport of particles in the volume
filled by the islands is turbulent and diffusive. Therefore, the
particle dynamics is modeled by the following Langevin equations:
\begin{equation}\label{lang}
\frac{d\mathbf{x}}{dt}={\boldsymbol\zeta}(t)\,\, ; \qquad \frac{du}{dt} =
W(\mathbf{x},t),
\end{equation}
with $\langle {\boldsymbol\zeta}(t)\rangle =0$ and $ \langle
\zeta_{i}(t) \, \zeta_{j}(t')\rangle \, =2\delta_{ij}\kappa_{T} \,
\delta(t-t')$, $\langle W(\mathbf{x},t)\rangle =0$ and $ \langle
W(0,0) \, W(\mathbf{x},t)\rangle  \, =C(\mathbf{x},t)$, $\kappa_{T}$
is the spatial diffusion coefficient. Here $C(\mathbf{x},t)$ is
the Eulerian correlation function associated with the
compression/expansion field $W(\mathbf{x},t)$ of the islands which
is supposed to be homogeneous and stationnary. The Eulerian correlation function
depends on three parameters that characterize the statistics of the (isotropic) compression/expansion field : the variance
$\langle W^{2}(\mathbf{x},t) \rangle=C(0,0)$, the correlation time $\tau_{c}$, which is the decay time of the Eulerian correlation and the
correlation length $\lambda_{c}$, which is the decay length. So the particles have probability to stay within the island
or escape. As noted \cite{2006Natur.443..553D}, the gyration radius of the particle increases near the separatrix, which in turn increases
the probability of a particle to escape the island. The Langevin equations (\ref{lang})
are doubly stochastic in the sense that both the position
$\mathbf{x}(t)$ of the particles and the compression field
$W(\mathbf{x},t)$ are stochastic processes.

With the spatio-temporal statistics of the compression being
specified via $C(\mathbf{x},t)$, the problem is to calculate the
diffusion coefficient in momentum space (when the latter exists) and
to determine the form of the distribution function. The diffusion
coefficient $D$ is related to the time integral of the Lagrangian
correlation function\cite{taylor}, viz.
\begin{equation}\label{taylor_result}
D=\int _{0}^{\infty} dt C_{L}(t),
\end{equation}
where the Lagrangian correlation function $C_{L}(t)$ is defined via
\begin{equation}\label{lag_corr}
C_{L}(t)=\langle W(0,0) \, W(\mathbf{x}(t),t)\rangle ,
\end{equation}
where $\mathbf{x}(t)$ is a solution of Eqs.(\ref{lang}). The exact
result (\ref{taylor_result}) is a simple consequence of the
definition $D=(1/2)d\langle u^{2}\rangle /dt$ combined with the second equation in
(\ref{lang}). Indeed, $\langle
u^{2}\rangle  \, = \int _{0}^{t}dt'\int _{0}^{t} dt''\langle
W(t')W(t'')\rangle  \, = \, 2\int _{0}^{t}dt' \, C_{L}(t') \,
(t-t')$ and letting $t\rightarrow \infty$ (when the integral
converges) gives Eq.(\ref{taylor_result}). It also follows from Eq.(\ref{taylor_result})
that the diffusion coefficient
in momentum-space can be expressed as
\begin{equation}\label{lag}
D=\tau_{L}\langle W^{2}(\mathbf{x},t)\rangle,
\end{equation}
where $\tau_{L}$ is the Lagrangian correlation time, i.e. the correlation time
of the compression/expansion field which is experienced by the particles along their trajectory.
The problem remains to connect Lagrangian and Eulerian statistics, i.e. to
determine the functional dependence of $\tau_{L}$ with $\tau_{c}$ and $\lambda_{c}$.
To this purpose, let us write the Lagrangian correlation
function~(\ref{lag_corr}) in the equivalent form
\begin{equation}\label{cldef}
C_{L}(t)=\int d\mathbf{x}\langle W(0,0) \, W(\mathbf{x},t) \, \delta
[\mathbf{x}-\mathbf{x}(t)]\rangle .
\end{equation}
A relation between the Lagrangian correlation $C_{L}(t)$ and the
Eulerian correlation $C(\mathbf{x},t)$ is obtained by invoking a
procedure due to Corrsin \cite{cor,tau} in which $\mathbf{x}(t)$ is
replaced by its statistical average, so that we may replace $\delta
[\mathbf{x}-\mathbf{x}(t)]$ in equation~(\ref{cldef}) by $\langle
\delta [\mathbf{x}-\mathbf{x}(t)]\rangle $.
 This leads to the factorization $
C_{L}(t)=\int d\mathbf{x}\langle W(0,0)W(\mathbf{x},t)\rangle  \, \langle \delta
[\mathbf{x}-\mathbf{x}(t)]\rangle $. Hence, an expression for the diffusion
coefficient in momentum space is found which is given by
\begin{equation}\label{aa}
D=\int_{0}^{\infty} dt \int d\mathbf{x} \, C(\mathbf{x},t) \
P(\mathbf{x},t) \,\,\, ,
\end{equation}
where $P(\mathbf{x},t)\equiv \langle
\delta[\mathbf{x}-\mathbf{x}(t)]\rangle $ is the conditional
probability for a particle to be under the influence of a magnetic
island located at the position $\mathbf{x}$ at time $t$ provided
that this particle was at $\mathbf{x}=0$ at $t=0$. Equation
(\ref{aa}) shows that $D$ is the
integral of the product of two quantities: $C(\mathbf{x},t)$, the
Eulerian correlation function, which characterizes the statistical
properties of the compression field $W(\mathbf{x},t)$, and the
probability function $P(\mathbf{x},t)$, describing the spatial
transport of particles in the acceleration region. Here,
$P(\mathbf{x},t)$ is the solution of a standard diffusion equation
with diffusion constant $\kappa_{T}$, i.e.
\begin{equation}\label{p}
P(\mathbf{x},t)=\frac{1}{(4\pi \kappa_{T}t)^{3/2}}\exp(-|\mathbf {x}|^{2}/4\kappa_{T}t),
\end{equation}
 but the procedure can be
generalized to more complex transport models.

\begin{figure}
\begin{center}
\includegraphics[width=0.5\textwidth]{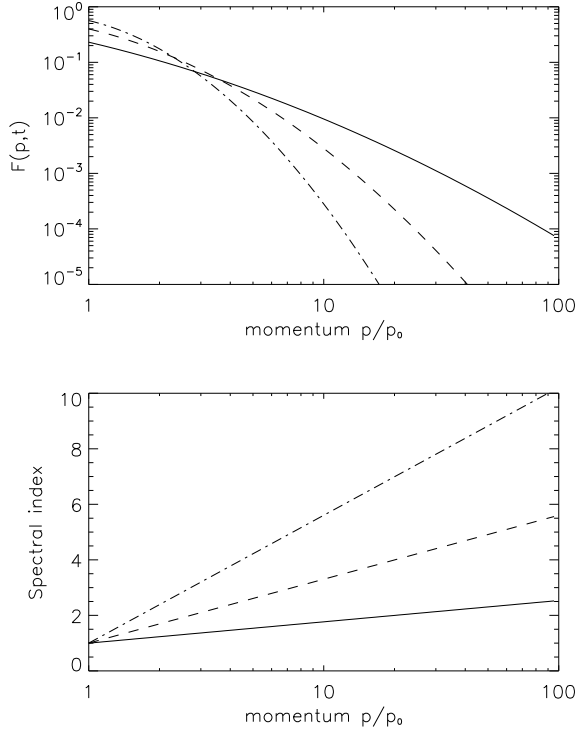}
\end{center}
\caption{Particle distribution function (top panel) and the spectral
index of the distribution (bottom panel). The solutions of Eq.(\ref{diffu})
for $Dt=3$ (solid line), $Dt=1$ (dashed line), and $Dt=0.5$ (dash-dotted line).
All distributions are normalised so that $\int F_{0}(p,t)dp=1$.}
 \label{fig1}
\end{figure}
Let us further take the illustrative example of
an isotropic correlation function of the form given by
\begin{equation}\label{c}
C(\mathbf{x},t)=\langle W^{2}(\mathbf{x},t)\rangle \exp (-|\mathbf{x}|^{2}/\lambda^{2}_{c}-t/\tau_{c}).
\end{equation}
From Eq.(\ref{aa}), we obtain that
\begin{equation}
D=\langle W^{2}(\mathbf{x},t)\rangle \int _{0}^{\infty} dt \exp (-t/\tau_{c})(1+\frac{4\kappa_{T}\tau_{c}}{\lambda_{c}^{2}})^{-3/2}.
\end{equation}
Therefore, in the weak
spatial diffusion limit, where $\kappa_{T} \ll
\lambda_{c}^2/4\tau_{c}$, the momemtum diffusion coefficient is given by
\begin{equation}
D\sim \tau_{c}\langle W^{2}(\mathbf{x},t) \rangle.
\end{equation}
This is the case already given by Eq.(\ref{con}) corresponding to Eq.(\ref{lag}) with $\tau_{L}\sim\tau_{c}$.
However, in the
opposite, strong spatial diffusion limit, where
$\kappa_{T} \gg \lambda_{c}^2/4\tau_{c}$, then
\begin{equation}
D\sim \frac{\lambda_{c}^{2}}{2\kappa_{T}}\langle W^{2}(\mathbf{x},t)\rangle,
\end{equation}
corresponding to the Lagrangian correlation time being
of the order of the spatial transport time scale, i.e.
$\tau_{L}\sim \lambda_{c}^{2}/\kappa_{T}$.  In this strong spatial
diffusion limit, the stochastic acceleration efficiency is governed
both by the Eulerian properties of the compression field and the
Lagrangian properties of the particles.

Let us notice that the
diffusion coefficient in momentum space may also be expressed as
$D=\int \int d\mathbf{k} \, d\omega \, S(\mathbf{k},\omega) \,
\kappa_{T} k^{2}/[\omega^{2}+(\kappa_{T} k^{2})^{2}]$, where $S(\mathbf{k},\omega)$ is the spectrum of
$W(\mathbf{x},t)$, i.e. the Fourier transform of the correlation function $C(\mathbf{x},t)$.
It can be clearly seen
from this expression for $D$ that the integral may
diverge for scale-free
power-law spectra such as $S(k,\omega)\propto k^{-q} \, \delta (\omega)$. This is the signal that the turbulent acceleration process
cannot be described as a standard diffusion in $u-$space as in
Eq.(\ref{diffu}). This situation has been dubbed Fermi acceleration of
fractional order in \cite{2008ApJ...687L.111B,2012ApJ...754..103B}. Here, we focus on the
second-order process with $D$ finite.

The statistical effect discussed above can be felt also in addition
to the systematic energy change. Indeed, when both the mean and the
variance of the compression are finite, the first and second-order
Fermi processes operate together. In this case, $F_{0}(u,t)$ obeys
an advection-diffusion equation in velocity space,
\begin{equation}\label{addif}
\frac{\partial F_{0}(u,t)}{\partial t}+a_{1}\frac{\partial}{\partial
u}F_{0}(u,t)=a_{2}\frac{\partial^{2} }{\partial u^{2} }F_{0}(u,t),
\end{equation}
where $u=\ln (p/p_{0})$ and where the coefficients of systematic and
stochastic acceleration are given by
\begin{equation}
a_{1}=\langle W(\mathbf{x},t)\rangle  ;
a_{2}=\tau_{L}\langle (W(\mathbf{x},t)-\langle W(\mathbf{x},t)\rangle )^{2}\rangle ,
\end{equation}
respectively. When the islands contract on average, the distribution
function $F_{0}(u,t)$ shifts toward large $u$ at a rate given by
$a_{1}$ while the variance of $F_{0}(u,t)$ grows at a rate given by
$a_{2}$ and the stochastic component to the acceleration process
dominates the systematic part when $a_{2}\gg a_{1}$. The
time-dependent solution $F_{0}(u,t)$ of the advection-diffusion
equation (\ref{addif}) is the normal distribution in the variable
$u-a_{1}t$.

Although the time-dependent solution is not a power
law, but only asymptotically at $t\rightarrow \infty$, the
characteristic solutions and spectral indices
\begin{equation}\label{sp_ind}
-\frac{d\ln F_{0}(p,t)}{d\ln p}=1+\frac{\ln p/p_0}{Dt},
\end{equation}
are found for a few values of $Dt$ and are presented in Figure \ref{fig1}.
The values appear to be similar to those obtained  in numerical
simulations e.g. \cite{2010ApJ...709..963D,2010ApJ...714..915O}
and closer to the observed values in solar flares \cite{2011SSRv..159..107H}
than for example in \cite{2006PhRvL..96o1102O}.

In summary, we show that both the first and second-order Fermi
acceleration process can operate together to increase the particle
energy when the acceleration region consists of an ensemble of
contracting islands. In the case when islands are both contracting
and expanding with zero mean effect, only the second-order
acceleration process operates. However, even when contraction is
dominant, the second order effect can be substantial. The stochastic
component to the acceleration corresponds to a non-resonant
mechanism according to the classification scheme established in
\cite{2012ApJ...754..103B}. It involves the turbulent transport properties of
the particles in the acceleration region and becomes more efficient
for higher levels of variance in the compression rate.

\begin{acknowledgments}
This work is supported by a STFC rolling grant. Financial
support by the European Commission through the "Radiosun" (PEOPLE-2011-IRSES-295272)
and HESPE (FP7-SPACE-2010-263086) is gratefully acknowledged.
\end{acknowledgments}

\bibliography{file1}

\end{document}